\documentclass[10pt,conference]{IEEEtran}
\IEEEoverridecommandlockouts

\usepackage{hyperref}
\usepackage{xspace}

\usepackage{url}
\usepackage{amsmath,amssymb,amsfonts}
\usepackage{amsthm}
\usepackage{subcaption}
\usepackage{textcomp}
\usepackage{enumitem}

\usepackage{multicol, multirow}
\usepackage{tikz}
\usepackage{comment}
\usepackage{pbalance}
\usepackage{pgfplots}
\usepackage{caption}

\usepackage{relsize}
\usepackage{wrapfig}
\usepackage[]{mdframed}
\usepackage[square,sort,comma,numbers]{natbib}
\usepackage{booktabs}
\usepackage{placeins}
\usepackage{adjustbox}

\usepackage[frozencache,cachedir=.]{minted}
\usepackage{algpseudocode}
\usepackage{tabularx}

\usepackage[linesnumbered,ruled,vlined]{algorithm2e}
\usepgfplotslibrary{groupplots}

\definecolor{darkgreen}{rgb}{0.0, 0.5, 0.0}

\usepackage{listings}
\usepackage{xcolor}
\lstdefinestyle{GoDemo}{
  language=Go,
  basicstyle=\small\ttfamily,
  keywordstyle=\bfseries,
  commentstyle=\itshape\color{gray},
  stringstyle=\color{teal},
  breaklines=true,
  breakatwhitespace=true,
  showstringspaces=false,
  frame=none,             
  captionpos=b,
  numbers=left,           
  numberstyle=\tiny,      
  numbersep=5pt,          
  stepnumber=1            
}

\AtBeginDocument{%
  }

\begin{document}

\title{FlakyGuard: Automatically Fixing Flaky Tests at Industry Scale}
\author{
\IEEEauthorblockN{Chengpeng Li\IEEEauthorrefmark{1}, 
Farnaz Behrang\IEEEauthorrefmark{2}, 
August Shi\IEEEauthorrefmark{1}, 
Peng Liu\IEEEauthorrefmark{2}}
\IEEEauthorblockA{\IEEEauthorrefmark{1}ECE Department, The University of Texas at Austin, USA\
\{{chengpengli, august}\}@utexas.edu}
\IEEEauthorblockA{\IEEEauthorrefmark{2}Uber Technologies, USA\
\{{behrang, peng3141}\}@uber.com}
}
\newtheorem{fact}{Fact}

\newcommand{\milind}[1]{{\color{red} Milind: #1}}
\newcommand{\chris}[1]{{\color{orange} Chris: #1}}
\newcommand{\farnaz}[1]{{\color{blue} Farnaz: #1}}
\newcommand{\peng}[1]{{\color{darkgreen} Peng: #1}}
\newcommand{\vlad}[1]{{\color{cyan} Vlad: #1}}
\newcommand{\chengpeng}[1]{{\color{purple} Chengpeng: #1}}
\newcommand{\drfix}{\textsc{Dr.Fix}}
\newcommand{\flakyguard}{\textsc{FlakyGuard}}

\newcommand{\UTAffiliation}{\affiliation{
    \department{Electrical and Computer Engineering}
    \institution{The University of Texas at Austin}
    \city{Austin}
    \state{TX}
    \country{USA}
    \postcode{78705}
}}

\newcommand{\UberAffiliation}{\affiliation{
    \department{Electrical and Computer Engineering}
    \institution{Uber Technologies}
    \country{USA}
}}

\newcommand{\red}[1]{{\color{red}#1}}
\newcommand{\Fix}[1]{\red{#1}}

\definecolor{impacted}{rgb}{0.7,0.7,0.7}
\definecolor{gitdel}{RGB}{255,236,236}
\definecolor{gitdelfocused}{RGB}{248,203,203}
\definecolor{gitadd}{RGB}{234,255,234}
\definecolor{gitaddfocused}{RGB}{166,235,166}

\makeatletter
\newenvironment{btHighlight}[1][]
{\begingroup\tikzset{bt@Highlight@par/.style={#1}}\begin{lrbox}{\@tempboxa}}
{\end{lrbox}\bt@HL@box[bt@Highlight@par]{\@tempboxa}{impacted}\endgroup}
\newcommand\btHL[1][]{%
  \begin{btHighlight}[#1]\bgroup\aftergroup\bt@HL@endenv%
}
\def\bt@HL@endenv{%
  \end{btHighlight}%
  \egroup
}
\newcommand{\bt@HL@box}[3][]{%
  \tikz[#1]{%
    \pgfpathrectangle{\pgfpoint{1pt}{0pt}}{\pgfpoint{\wd #2}{\ht #2}}%
    \pgfusepath{use as bounding box}%
    \node[anchor=base west, fill=#3, outer sep=0pt,inner xsep=0pt, inner ysep=-0.5pt, #1]{\raisebox{1pt}{\strut}\strut\usebox{#2}};
  }%
}
\makeatother

\newcommand{\XComment}[1]{}
\newcommand{\XSpace}[1]{}
\newcommand{\CodeIn}[1]{{\texttt{#1}}}

\newcommand{\etal}{et al.}

\newcommand{\company}{Uber}
\newcommand{\SmartBFS}{\textit{FlakyGuard}}
\newcommand{\NOPROD}{\textit{FlakyDoctor}}
\newcommand{\NAIVE}{\textit{UUT}}
\newcommand{\BFSRepograph}{\textit{BFSRepoGraph}}
\newcommand{\BFSDCG}{\textit{BFSDCG}}
\newcommand{\Agentless}{\textit{Agentless+RepoGraph}}
\newcommand{\Repograph}{\textit{RepoGraph}}
\newcommand{\AutoCodeRover}{\textit{AutoCodeRover}}

\newcommand{\gptmodel}{ChatGPT o1}
\newcommand{\claudemodel}{Claude 3.7 Sonnet}
\newcommand{\noprodtime}{1680.33}
\newcommand{\naivetime}{1462.50}
\newcommand{\bfsrepographtime}{1517.84}
\newcommand{\bfsdcgtime}{1952.64}
\newcommand{\smartbfstime}{1978.73}
\newcommand{\claudetime}{1829.27}
\newcommand{\agentlesstime}{1528.60}
\newcommand{\autocoderovertime}{1799.74}

\newcommand{\reproductionRate}{50.5\%}
\newcommand{\rejectChangeLogic}{26.9\%}
\newcommand{\rejectWrong}{19.2\%}
\newcommand{\rejectLong}{7.7\%}

\newcommand{\numFixedSmartBFS}{194}
\newcommand{\numFixedNoProd}{134}
\newcommand{\numFixedNaive}{148}
\newcommand{\numFixedBFSRepograph}{149}
\newcommand{\numFixedBFSDCG}{161}
\newcommand{\numFixedSimplification}{160}
\newcommand{\numFixedAgentless}{159}
\newcommand{\numFixedAutoCodeRover}{158}
\newcommand{\numFixedBFSClaude}{173}
\newcommand{\numFixedBFSGPTIV}{97}

\newcommand{\numFixedBySBFSNotAgent}{35}
\newcommand{\numFixedBySBFSNotAutocoderover}{36}
\newcommand{\numFixedByGPTNotClaude}{50}
\newcommand{\numFixedByClaudeNotGPT}{29}
\newcommand{\percentageFixedSimplification}{54.24\%}
\newcommand{\percentageFixedSmartBFS}{65.76\%}

\newcommand{\RQOne}{How effective is \flakyguard{} at fixing flaky tests?}
\newcommand{\RQTwo}{What is the contribution of each component of the technique to the overall results?}
\newcommand{\RQThree}{How does \flakyguard{} compare with the state-of-the-art
approaches?}
\newcommand{\RQFour}{How much do the results change when utilizing a different model?}
\newcommand{\RQFive}{How do developers perceive the usefulness of \flakyguard{}?}

\newcommand{\monorepodevs}{$\sim$6000\xspace}
\newcommand{\noappxmonorepodevs}{6000\xspace}

\newcommand{\numdevs}{thousands\xspace}
\newcommand{\totaldrs}{\red{total data races?}\xspace}
\newcommand{\fixeddrs}{\red{fixed data races}\xspace}
\newcommand{\period}{three\xspace}
\newcommand{\timeperdr}{11 days\xspace}
\newcommand{\monorepoloc}{100 million\xspace}
\newcommand{\monoreposvs}{$\sim$3500\xspace}
\newcommand{\fixeddrrate}{86\%\xspace}
\newcommand{\numdrfixfixed}{193\xspace}
\newcommand{\numdrfixdevs}{a hundred\xspace}
\newcommand{\numcores}{$\sim$2 million\xspace}
\newcommand{\diffsperday}{$\sim$1500\xspace}
\newcommand{\mgo}[1]{\texttt{\small  #1}}

\newcommand{\mgotype}[1]{{\text{\relsize{0}{\tt\color{black}\textbf{#1}}}}}
\newcommand{\mgobuiltin}[1]{{\text{\relsize{0}{\tt\color{black}\textbf{#1}}}}}
\newcommand{\mcode}[1]{{\text{\tt{#1}}}}
\newcommand{\hb}{$HB$}
\newcommand{\drsfound}{$8000$}
\newcommand{\drsfixed}{$4000$}
\newcommand{\dailyraces}{$\textrm{5-15}$}
\newcommand{\rrt}{${\color{red}\blacktriangleright}$}
\newcommand{\grt}{${\color{green}\blacktriangleright}$}
\newcommand{\gplus}{\color{darkgreen}+}
\newcommand{\rminus}{{\color{red}-}}

\newcommand{\testrepeats}{$1000$}
\newcommand{\drfixperiod}{18-month}
\newcommand{\drfixperiodnohyphern}{18 month}

\newcommand{\cureatedbugs}{272}
\newcommand{\cureatedsupersetbugs}{747}

\newcommand{\RQfirstnumOfTests}{1115}
\newcommand{\RQfirstnumOfReproducibleTests}{798}
\newcommand{\RQfirstnumGeneratedFixes}{380}
\newcommand{\RQfirstnumAccepted}{197}
\newcommand{\RQfirstrejectInRight}{48.2\%}

\newcommand{\RQSecondnumOfReproducibleTests}{295}
\newcommand{\RQSecondnumOfTestTargets}{173}
\newcommand{\RQSecondnumOfTests}{559}
\newcommand{\RQSecondnumGeneratedFixes}{194}
\newcommand{\RQSecondnumAccepted}{60}


\newtheorem{Definition}{Definition}[section]
\newtheorem{Theorem}{Theorem}
\newtheorem{Remark}{Remark}
\newtheorem{Observation}{Observation}
\newtheorem{Rule}{Rule}
\newtheorem{Fact}{Fact}
\newtheorem{Lemma}{Lemma}
\newtheorem{Corollary}{Corollary}
\newtheorem{Lcorol}{Corollary}
\newtheorem{Example}{Example}

\lstdefinelanguage{CustomGo}{%
aboveskip=5pt,
belowskip=0pt,
lineskip= {-1.5pt},
language=go,                %
basicstyle=\ttfamily\scriptsize,       %
numbers=left,                   %
numberstyle=\tiny,      %
stepnumber=1,                   %
numbersep=2pt,                  %
backgroundcolor=\color{white},  %
showspaces=false,               %
stringstyle=\scriptsize,
identifierstyle=\scriptsize,
commentstyle=\scriptsize,
basicstyle=\scriptsize\ttfamily,
showstringspaces=false,         %
showtabs=false,                 %
frame=tb,                   %
tabsize=2,                      %
captionpos=b,                   %
breaklines=true,                %
breakatwhitespace=false,        %
title=\lstname,                 %
  sensitive,%
  morecomment=[s]{/*}{*/},%
  morecomment=[l]//,%
  morecomment=[f][{\btHL[fill=gitdel]}]-,
  morecomment=[f][{\btHL[fill=gitadd]}]+,
  morestring=[b]',%
  morestring=[b]",%
  morestring=[s]{`}{`},%
  morekeywords=[1]{break,case,const,continue,default,defer,%
      else,fallthrough,false,for,func,go,goto,if,import,iota,skip,%
      range,return,select,switch,true,type,nop,%
      var,then,while},%
  morekeywords=[3]{append,cap,close,complex,copy,delete,%
      len,make,new,panic,print,println,recover},%
  morekeywords=[2]{bool,map,byte,complex64,complex128,float32,float64,%
      int,int8,int16,int32,int64,rune,string,interface,struct,%
      uint,uint8,uint16,uint32,uint64,uintptr,chan,error,any},%
  keywordstyle=[1]{\bfseries\color{black}},
  keywordstyle=[2]{\bfseries\color{black}},
  keywordstyle=[3]{\bfseries\color{black}},
  commentstyle=\color{gray},
  backgroundcolor=\color{white},
  escapechar={@},
  numbersep=5pt,
  xleftmargin=10pt,
  numbers=left,
}%

\lstdefinelanguage{CustomPyton}{%
aboveskip=5pt,
belowskip=0pt,
lineskip= {-1.5pt},
language=python,                %
basicstyle=\scriptsize,       %
numbers=left,                   %
numberstyle=\tiny,      %
stepnumber=1,                   %
numbersep=2pt,                  %
backgroundcolor=\color{white},  %
showspaces=false,               %
stringstyle=\scriptsize,
identifierstyle=\scriptsize,
commentstyle=\scriptsize,
basicstyle=\scriptsize\ttfamily,
showstringspaces=false,         %
showtabs=false,                 %
frame=tb,                   %
tabsize=2,                      %
captionpos=b,                   %
breaklines=true,                %
breakatwhitespace=false,        %
title=\lstname,                 %
  sensitive,%
  morecomment=[s]{/*}{*/},%
  morecomment=[l]//,%
  morestring=[b]',%
  morestring=[b]",%
  morestring=[s]{`}{`},%
  commentstyle=\color{gray},
  backgroundcolor=\color{white},
  escapechar={@},
  numbersep=5pt,
  xleftmargin=10pt,
  numbers=left,
  morekeywords=[1]{if,for,def,range,retrun,},%
  keywordstyle=[1]{\bfseries\color{black}},
}%

\maketitle

\begin{abstract}
Flaky tests that non-deterministically pass or fail waste developer time and slow release cycles. While large language models (LLMs) show promise for automatically repairing flaky tests, existing approaches like FlakyDoctor fail in industrial settings due to the \emph{context problem}: providing either too little context (missing critical production code) or too much context (overwhelming the LLM with irrelevant information). We present \flakyguard{}, which addresses this problem by treating code as a graph structure and using selective graph exploration to find only the most relevant context.
Evaluation on real-world flaky tests from industrial repositories shows that \flakyguard{} repairs 47.6\% of reproducible flaky tests with 51.8\% of the fixes accepted by developers. Besides it  outperforms state-of-the-art approaches by at least 22\% in repair success rate. Developer surveys confirm that 100\% find \flakyguard{}'s root cause explanations useful.
\end{abstract}

\section{Introduction}
\label{sec:intro}

Flaky tests non-deterministically pass or fail when rerun on identical code, creating significant challenges in software development. When tests fail, developers cannot distinguish between those failures indicating real bugs or flakiness, forcing them to spend time investigating false alarms. In many industrial settings, teams may block code deployment until all test failures are resolved, slowing down release cycles. The prevalence of flaky tests wastes both developer time and computational resources~\cite{computeresource1,computeresource2}.

Recent advances in large language models (LLMs)~\cite{openaitr, claude, tracingthought}, particularly in code generation and reasoning capabilities, show great promise for automatically repairing flaky tests. Prior work has demonstrated this potential: FlakyDoctor~\cite{flakydoctor} successfully repaired more flaky tests than traditional non-LLM approaches~\cite{odrepair, ifixflakies, dexfix} in open-source projects\XSpace{, handling a wider range of flaky test types compared to the specific, targeted repairs of non-LLM-based methods}. Importantly, non-LLM approaches require prior tools to classify flaky test types before applying pattern-specific fixes, limiting their applicability when such classification is unavailable, as is often the case in industrial settings where only test failure information is readily accessible.

However, our experiments with FlakyDoctor in an industrial setting at \company{} revealed significant limitations, as it frequently failed to repair real-world flaky tests. Our analysis identified the primary cause as the \emph{context problem}, which manifests in two ways. On one hand, the LLM may receive \emph{too little} context: FlakyDoctor~\cite{flakydoctor} provides only the test code as context, causing the LLM to miss critical production code information necessary for proper reasoning and root cause analysis. On the other hand, the LLM may receive \emph{too much} context: including all production code from a service (e.g., medium-sized services at \company{} typically exceed 100K+ LOCs) overwhelms the LLM, diffusing its attention~\cite{vaswani2017attention}, and degrading performance on complex reasoning tasks such as root cause analysis.
Additionally, developers usually need clear explanations of root causes and repair rationale to accept proposed fixes.
Therefore, the goal is to provide the right balance of context that is neither too much nor too little, enabling the LLM to not only repair the flaky test but also explain the root cause to developers.


Recent LLM-based approaches~\cite{agentless, autocoderover} designed for other software engineering tasks have proposed context collection methods that can be adapted to flaky test repair with slight modifications (Section~\ref{sec:eval}). While directly applying these context collection techniques to FlakyDoctor~\cite{flakydoctor} improves repair rates, significant limitations remain in our domain. Specifically, Agentless~\cite{agentless} and AutoCodeRover~\cite{autocoderover} construct context by treating the codebase as a flat text corpus, missing opportunities to leverage code structure for more targeted context selection.

Treating the codebase as a graph structure enables more effective context collection by using the call graph among functions to guide the search, entirely avoiding irrelevant functions through structured graph traversal. However, analyzing the codebase as a graph requires a novel \emph{graph search} design, as naive BFS/DFS approaches would include the entire graph, overwhelming the LLM. Existing graph-based work RepoGraph~\cite{repograph} uses ego-graphs~\cite{egograph} that include $k$-hop nodes around central nodes to avoid exponentially growing context, but this approach is severely limited by depth constraints. Our preliminary study revealed that critical repair information often resides in leaf nodes deeply embedded in the call graph.
To address these limitations, we develop a \emph{selective graph exploration} strategy that identifies and traverses only the most relevant paths in the call graph. Rather than being constrained by fixed depth limits, our approach uses the LLM itself to guide the graph exploration process, enabling it to reach critical information regardless of its depth while maintaining manageable context size.

To further boost the effectiveness of the graph search, we also use a dynamic call graph (DCG) collected from test runs. A DCG captures the precise scope of functions actually executed at runtime, enabling our search to follow only the executed paths. In contrast, static call graphs may include functions that are never called by the specific test case. In addition, static call graphs struggle with overloaded functions, anonymous functions, and reflection-based calls, creating imprecise linkages that can misguide the LLM to either include irrelevant functions or miss relevant ones. The use of a DCG resolves these issues.


We develop \flakyguard{}, which addresses the context problem through LLM-guided exploration of dynamic call graphs. Our approach constructs dynamic call graphs from test execution traces, then employs an iterative LLM-based selection process to expand only those paths deemed relevant for understanding flaky behavior. This selective traversal enables \flakyguard{} to reach critical information at arbitrary depths while maintaining manageable context sizes for effective repair and clear root cause explanations.

Additionally, Go codebases in industrial settings predominantly use table-driven testing~\cite{tabledriventests}, where test functions contain multiple similar test cases organized in tables. This paradigm creates challenges for existing approaches~\cite{flakydoctor, agentless, repograph, autocoderover}, which frequently analyze the wrong test cases or encounter patching failures due to ambiguity in search-and-replacement operations~\cite{agentless}. We develop AST-based \emph{test simplification} and \emph{patch transplantation} techniques (Section~\ref{sec:simplification_patching}) to address these challenges and apply them to all baseline approaches to ensure they produce meaningful results.

We evaluate \flakyguard{} on real-world flaky tests of various types from industrial repositories at \company{}.
Over six months of deployment, \flakyguard{} produced fixes for 47.6\% of reproducible flaky tests, with 51.8\% of generated fixes accepted by developers.
Compared to state-of-the-art LLM-based repair methods~\cite{flakydoctor, agentless, repograph, autocoderover}, \flakyguard{} outperforms them by at least 22\% in repair success rate. In anonymous developer survey responses, 100\% of the developers found \flakyguard{}'s root cause explanations useful. We also conduct comprehensive ablation studies, case studies, and breakdown analyses to address key research questions.

Beyond the experimental validation, \flakyguard{} has been deployed as a fully autonomous system at \company{}, integrating with the existing ticketing system to automatically process flaky test reports and deliver fixes to developers daily.

In this paper, we make the following contributions:
\begin{itemize}
    \item We identify and characterize the \emph{context problem} in LLM-based flaky test repair, where existing approaches provide either too little or too much context.
    \item We develop \flakyguard{}, a novel approach that uses LLM-guided exploration of dynamic call graphs to reach critical information at arbitrary depths while maintaining manageable context sizes.
    \item We conduct a large-scale evaluation in industrial settings, showing that \flakyguard{} outperforms state-of-the-art approaches and produces useful root cause explanations.
\end{itemize}

\section{Running Example}
\label{sec:running}


\begin{figure}
    \begin{lstlisting}[frame=single, basicstyle=\ttfamily, xleftmargin=0.1em, xrightmargin=0.1em, breaklines=true, linewidth=1\linewidth, language=CustomGo, label={casestudy:runningexample}, caption=Flaky Test that needs the Deep Calling Context, captionpos=b]
// backend_test.go
...
  t.Run("validation and call UpdateInfo", func(t *testing.T) {
    ...
@\gplus@   @\textcolor{darkgreen}{var wg sync.WaitGroup}@
@\gplus@   @\textcolor{darkgreen}{wg.Add(1)}@
@\rminus@   @\textcolor{red}{doc.Let().UpdateInfo().Return(nil)}@
@\gplus@   @\textcolor{darkgreen}{doc.Let().UpdateInfo().Run(func(ctx context.Context, data *entity.Info) \{ }@
@\label{line:running:waitdone}@@\gplus@   @\textcolor{darkgreen}{ \ \ \ \    defer wg.Done()}@
@\gplus@   @\textcolor{darkgreen}{\}).Return(nil)}@

    ...
    err := h.AddProgram(context.Background(), r)
    ...
@\label{line:running:wait}@@\gplus@   @\textcolor{darkgreen}{wg.Wait()}@
    assert.NoError(t, err)
  })

// deep call chain leading to this is omitted.
// validator.go
func ValidateIdentity(ctx context.Context, program *entity.Program, c *controller) map[string]string {
  ...
  data, err := c.db.GetInfo(ctx, _a, email)
@\label{line:running:updateinfo}@  go c.db.UpdateInfo(ctx, data)
  ...
}
...

    \end{lstlisting}
\end{figure}

We illustrate the challenge of repairing flaky tests using the example in Listing~\ref{casestudy:runningexample}.
The test in the source file backend\_test.go non-deterministically fails with the error ``Not all calls expected by the mock for UpdateInfo were made''. This error originates from the mock library, which verifies that all expected mock calls have been executed when the test completes, but it is not a standard assertion failure. Notably, this error provides no stacktrace, and even when stacktraces are available, they typically only show what happened rather than why the error occurred.

The function {\tt UpdateInfo} is invoked through a deep call chain: {\tt AddProgram} (backend.go) $\rightarrow$ {\tt AddProgram} (controller.go) $\rightarrow$ {\tt ValidateIdentity} (validator.go) $\rightarrow$ {\tt UpdateInfo}, with each function residing in a different source file. Critically, {\tt UpdateInfo} is invoked as a goroutine without synchronization with the test execution. The goroutine may not complete when the test finishes, especially in the heavily loaded CI environments, thereby causing the mock call to be missed and triggering the error.

The fix produced by \flakyguard{} adds a {\tt WaitGroup} to synchronize the test with the goroutine: the test waits (line~\ref{line:running:wait}) until the goroutine signals completion (line~\ref{line:running:waitdone}) after making the mock call (line~\ref{line:running:updateinfo}).
Discovering this fix automatically is challenging because root cause analysis requires the LLM to trace through the call chain across multiple files to identify that {\tt UpdateInfo} runs in an unsynchronized goroutine. This scenario exemplifies the fundamental \emph{context problem}: without sufficient context, the LLM fails to understand why the test is flaky, yet too much context overwhelms the LLM with irrelevant information. We examine how different context collection strategies perform on this example.

\noindent
\textbf{No Production Code} Without production code context, the LLM initially tries to remove the mock entirely, which fails to pass the validations, then settles on relaxing the mock expectation with {\tt .MaxTimes(1)}. While this eliminates flakiness, it weakens the test by allowing the mock call to be skipped, which is precisely what the test should detect as a bug.


\noindent
\textbf{Text Search--based Approaches} We tested Agentless~\cite{agentless} and AutoCodeRover~\cite{autocoderover}. Agentless uses hierarchical search (files by names, then functions by signatures, then code blocks), while AutoCodeRover provides APIs like {\tt search\_method\_in\_file} and lets the LLM decide which APIs to call.

Both failed to repair the flaky test, collecting irrelevant context while missing the critical {\tt ValidateIdentity} function. These agents rely on natural language understanding to guide search, but {\tt ValidateIdentity} provides no hint that it calls {\tt UpdateInfo}. With 35+ functions per file and nearly 50 calls in {\tt AddProgram} alone—including five functions with the ``Validate'' prefix—the noise overwhelms context selection.

\noindent
\textbf{Graph based Approaches} We attempted using RepoGraph~\cite{repograph, repograph_github}, which extends Agentless~\cite{agentless} by building static call graphs and retrieving k-hop ego-graphs around relevant functions. RepoGraph keeps $k$ small (1-2 hops) to avoid exponential context growth. However, it fails to help in this example, because RepoGraph avoids deep traversal to leaf nodes such as {\tt ValidateIdentity}, precisely where the root cause resides. Moreover, its static call graph includes unexecuted branches, adding noise that confuses the LLM.

\noindent
\textbf{Our Work} \flakyguard{} addresses these limitations using LLM-guided selective graph search on top of dynamic call graphs.
First, the dynamic call graph contains only 40\% of the nodes compared to static call graphs by pruning unexecuted branches and precisely resolving overloaded functions. This filtering bounds the search space to actually executed function chains, reducing noise.
Second, our selective graph search can traverse deeply to leaf nodes like {\tt ValidateIdentity}, enabling the LLM to generate not only correct fixes but also reasonable root cause explanations.

\section{\flakyguard{} Methodology}
\label{sec:method}

\begin{figure}[t]
    \centering
    \includegraphics[width=1.0\linewidth]{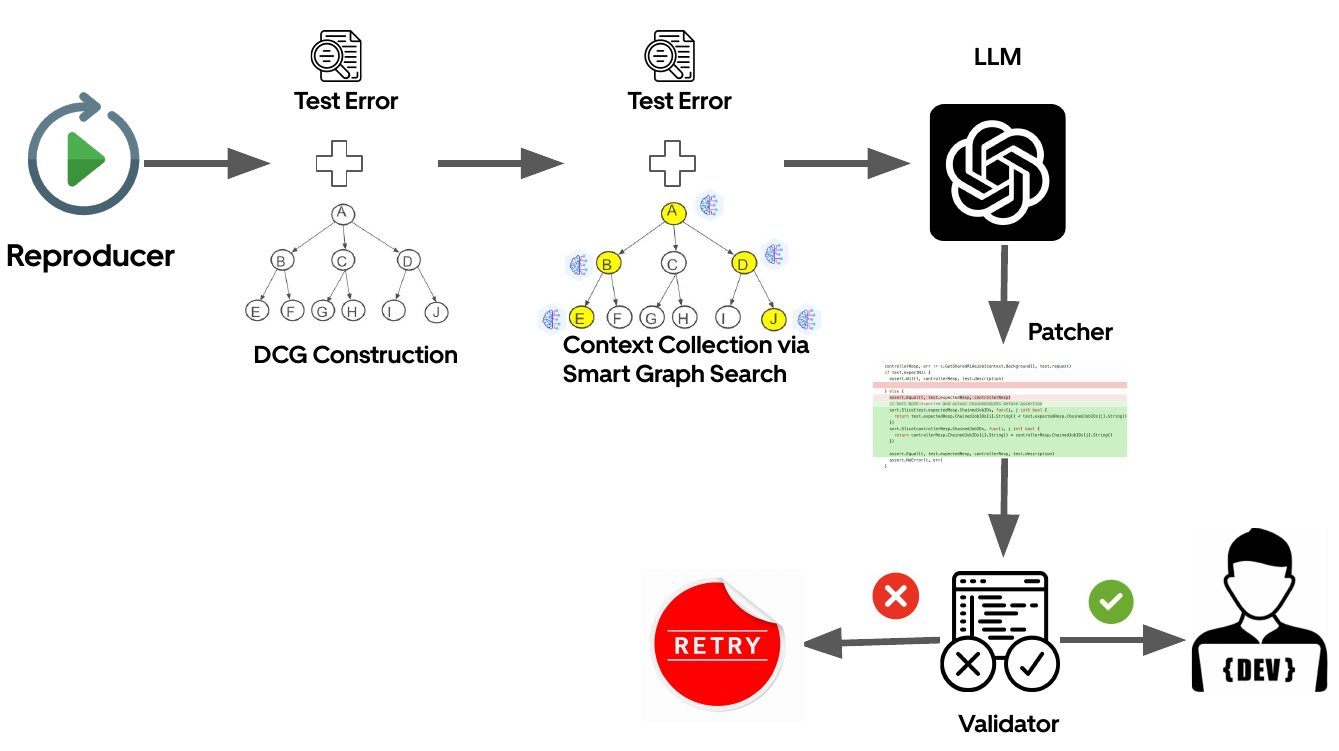}
    \caption{Overview of the \flakyguard{} Workflow.}
    \label{fig:overview}
\end{figure}

Figure~\ref{fig:overview} shows the high-level overview. Given a flaky test reported by our ticketing system, \flakyguard{} reproduces flaky test failures, builds a dynamic call graph (Section~\ref{sec:call_graph}), gathers context via LLM-guided graph traversal (Section~\ref{sec:graph_traversal}), and uses prompts to request the fixes from the LLM. After receiving a suggested fix, \flakyguard{} applies and validates it, and automatically submits it for developer review if validation succeeds. Otherwise, it enters the iterative fixing loop (Section~\ref{sec:main_fixing}). Lastly, Section~\ref{sec:simplification_patching} describes \flakyguard{}'s test simplification and patch transplantation techniques, which are generally applicable to existing approaches~\cite{agentless,autocoderover,repograph} for addressing key challenges in table-driven testing.

Before we dive into each component, we first introduce the inputs of \flakyguard{} and how it reproduces the flaky test failures to collect the error details. These steps are crucial, because the error details are not always available in the ticketing system, or they may be stale due to recent code changes.

\textbf{Inputs of \flakyguard{}} \flakyguard{} receives the flaky tests reported by our ticketing system (out of scope), e.g., in the conventional \textit{Bazel}  form of \url{test_target/test_func/test_case}, where \texttt{test\_target} consists of multiple test functions, and each \texttt{test\_func} contains multiple test cases following the predominantly used table-driven testing pattern~\cite{tabledriventests}.

\textbf{Failure reproduction.} To collect failure information, we execute the test case $N$ times (default: 1000). If no failures occur, indicating potential interference from other test cases, we execute the entire test target $N$ times again and filter results for the targeted test case. The workflow terminates if reproduction fails.

\textbf{Information extraction.} From reproduced failures, we extract: error message, stack trace, assertion file path and line number, and test function file path. These paths may differ when tests invoke external assertion utilities. We also extract the assertion statement using the file path and line number. We use regular expressions to parse this information, using LLM fallback for rare corner-cases.

\textbf{Prompting} \flakyguard{} uses a general prompt design not customized for specific flakiness types. The prompt combines a system message, extracted failure information, and context from graph traversal (Sections~\ref{sec:call_graph} and~\ref{sec:graph_traversal}). In the system prompt, we use
``you are an expert in fixing flaky tests'', and we specify that the LLM should only make changes to the test code, not the production code, which is beyond the scope of \flakyguard{}.

\subsection{Construction of Dynamic Call Graphs}~\label{sec:call_graph}
To construct the dynamic call graph, \flakyguard{} instruments the source code at compile time to inject the logging code~\footnote{We follow techniques from bazel/rules\_go~\cite{rulesgo4102}.}, which collects call edges at runtime.

The instrumentation modifies the original source code. To ensure the call graph accurately reflects execution, we collect the call graphs specific to the failed runs. The instrumentation is removed before requesting fixes from the LLM, with fixes later transplanted back to the original source (Section~\ref{sec:simplification_patching}).

At runtime, the logging code collects call edges in the format \texttt{MethodEntry: file, line, method Caller: file, line, method}.
It also dedups to log only the new edges, and it flushes the file upon every new edge to ensure all edges are written before test completion. With the collected logs, \flakyguard{} builds a dynamic call graph by mapping each  method to a unique AST node using the file path and line number, with each file mapped to a unique AST tree.

\textbf{Encoding Function Calls via Goroutines} Consider this minimal example of commonly used goroutine calls: \texttt{func A()\{go B()\}}. The call edge from \texttt{A} to \texttt{B} is not captured in the log because \texttt{B} is launched in a new goroutine, and the runtime stack does not include the enclosing function \texttt{A}.

We address this issue by analyzing the AST of the function \texttt{A} that launches the goroutine, identifying the functions being called, e.g., \texttt{B}. We then link the nodes of \texttt{A}  and \texttt{B}.

This name-based linkage is not perfect, as it may introduce spurious edges due to interface call resolution ambiguity. Also, if goroutines originate from external libraries without source code, the call edge is lost, and those functions like \texttt{B} become root nodes in the graph traversal. We found those issues do not impact \flakyguard{} much in practice, as only around 0.2\% of code starts goroutines from external libraries.

\subsection{Context Collection with LLM-guided Graph Traversal}~\label{sec:graph_traversal}
Algorithm~\ref{alg:smartbfs} performs a breadth-first search (BFS) where a LLM selects which nodes to explore rather than exploring all neighbors. Starting from root nodes, the algorithm queries the LLM at each step to select at most $k$ children per node based on their relevance to the flaky test error, continuing to depth $d$ (configurable, set to infinite by default). This step produces a list of selected nodes in BFS order.

As a final post-processing step, the LLM selects up to $F$ most relevant functions from this result to reduce context size, in addition to the root nodes that are always included. Unlike existing approaches that use depth-limited search~\cite{repograph}, this method is more effective since it identifies globally the relevant functions rather than imposing arbitrary depth constraints.

\begin{algorithm}
\caption{Smart Graph Traversal with GenAI}
\label{alg:smartbfs}
\KwIn{Call graph $G$, depth limit $d$, children per node $k$}
\KwOut{List of selected nodes $L$}

\SetKwFunction{CollectContext}{CollectContext}
\SetKwProg{Fn}{Function}{:}{}
\Fn{\CollectContext{$G$, $d$, $k$}}{
    $L \gets \text{GetRootNodes(G)}$\;

    $Q \gets$ queue of tuples $(n, 0)$ for nodes $n \in L$ \tcp*{each tuple: (node, depth)}

    \While{$Q$ is not empty}{
        $(n, h) \gets Q.\text{pop}()$\;

        \If{$h \geq d$}{
            \textbf{continue}\;
        }

        $C \gets \text{GetChildren}(n)$\;

        $S \gets \text{GenAI\_Select}(C, k)$\;

        \ForEach{$s \in S$}{
            $L \gets L \cup \{s\}$\;

            $Q.\text{append}(s, h + 1)$\;
        }
    }

    \Return $L$\;
}

\end{algorithm}

\subsection{Fixing Loops}\label{sec:main_fixing}


At a high level, \flakyguard{} operates through three nested loops: the outer loop ($M$ iterations) collects new contexts, the middle loop ($P$ iterations) generates high-level thoughts by prompting the LLM, and the inner loop ($N$ iterations) produces, applies, and validates fixes for each thought.
Each loop uses progressively enriched prompts: the outer loop uses the flakiness information, the middle loop adds context from graph traversal, and the inner loop incorporates the proposed thought.
We explain the configurable parameters in Section~\ref{sec:setup}.

In the middle loop, each thought contains root cause category, root cause explanation, and fixing plan. Following prior work~\cite{flakyfix}, we provide the LLM with summarized categories from past fixes while allowing the LLM to propose new categories. Failed thoughts are recorded to help the LLM avoid ineffective strategies, following the Reflection paradigm~\cite{reflection}.

In the inner loop, \flakyguard{} applies and validates LLM-suggested fixes. Fix application is detailed in Section~\ref{sec:simplification_patching}. Validation consists of build validation (ensuring compilation) and test validation (rerunning tests the same number of times as failure reproduction). For compilation failures, we attempt best-effort repair by providing the LLM with the original code, the modified code, and the compilation errors. Test validation requires all runs to pass; any error or timeout triggers a revert and iterating to the next fix. \flakyguard{} sends the fix to developers for review upon successful validation.

\subsection{Test Simplification and Patch Transplantation}\label{sec:simplification_patching}

Following the table-driven testing practice~\cite{tabledriventests}, each test function contains multiple similar test cases organized into tables, which creates challenges for fixing: the LLM may focus on the wrong test case, or search-and-replacement patching may fail due to ambiguity. We develop AST-based \emph{test simplification} and \emph{patch transplantation} techniques to address these challenges.

Before flakiness reproduction, \flakyguard{} simplifies the test function $T_{\text{orig}}$ to keep only the targeted test case $t$ while removing all other sibling test cases, yielding a simplified test function $T_{\text{simp}}$. We disable simplification for tests that need interference with others to manifest. This simplification process traverses the AST to record the byte offsets of all changes in an editing tracker. To preserve offset validity during modification, the changes are applied in reverse order~\cite{rulesgo4102}. After simplification, some variables become unused, causing compilation errors. We handle these errors iteratively by commenting them out, adding annotations to them for later restoration.







We send $T_{\text{simp}}$ to the LLM to produce the fix $T_{\text{simp}}'$. The challenge is then transplanting these edits back to the full context of $T_{\text{orig}}$, yielding a patched test function $T'_{\text{orig}}$:

\[
\texttt{Patch}(T_{\text{simp}}, T_{\text{simp}}', T_{\text{orig}}) \to T_{\text{orig}}'
\]


We use a two-step AST-based transplantation approach. First, we extract the AST from $T_{\text{simp}}'$ and replace the enclosing test table with its counterpart from $T_{\text{orig}}$. This process preserves the fixes outside the test table, which is crucial because the LLM may suggest changes to the surrounding code, not just the test case itself.
Second, we reload the merged AST and precisely replace the specific test case node from $T_{\text{orig}}$ with the corresponding fixed node from $T_{\text{simp}}'$. This approach ensures that fixes are correctly propagated while maintaining the original table structure.




\section{Evaluation}
\label{sec:eval}

To evaluate the effectiveness and usefulness of \flakyguard{} in producing fixes for flaky tests, we conducted a series of experiments designed to answer the following research questions:

\begin{enumerate}
    \item \textbf{RQ1.} \RQOne{}
    \item \textbf{RQ2.} \RQTwo{}
    \item \textbf{RQ3.} \RQThree{}
    \item \textbf{RQ4.} \RQFour{}
    \item \textbf{RQ5.} \RQFive{}
\end{enumerate}

\subsection{Experiments Setup}
\label{sec:setup}

We use the Go monorepo and its real-world flaky tests in \company{} to conduct our evaluation. This monorepo consists of over 100 million lines of Go code across more than 100 distinct projects, developed by over 6000 engineers from 100+ teams spanning diverse domains including storage systems, machine learning applications, and distributed services. Our evaluation dataset provides comprehensive coverage of different types of flaky tests since it systematically captures every flaky test occurrence rather than being constrained to predetermined categories. We choose \gptmodel~\citep{o1} and \claudemodel~\cite{claude} as the LLM models, because they are the state-of-the-art models available at the time of this evaluation.

For our experiments, we set a time limit of 2 hours for each fixing process. We run each test 1000 times to collect a failing run. To increase the chance of detecting scheduling-related flaky failures, we enabled the race detection flag~\cite{rules_go_race_detector} to introduce randomness at scheduling points.

For the parameters described in Sections~\ref{sec:graph_traversal} and ~\ref{sec:main_fixing}, we control the number of attempts for collecting context $M$ to be 3, the number of high-level thoughts $P$ per context to be 2, and the number of fix attempts per thought $N$ to be 3. The depth limit $d$ is set to infinite (i.e., max integer). The number of functions $F$ that the postprocessing retains is maximally 5.



\subsection{RQ1: \RQOne{}} \label{subsec:rq1}

We analyzed a total of \RQfirstnumOfTests{} flaky tests that \flakyguard{} attempted to fix, over a period of six months. Among these flaky tests:

\begin{itemize}
  \item \flakyguard{} reproduced $\mathbf{71.6\%}$ of the tests (i.e., $\mathbf{\RQfirstnumOfReproducibleTests{}}$ tests),
  \item \flakyguard{} produced fixes for $\mathbf{47.6\%}$ (i.e., $\mathbf{\RQfirstnumGeneratedFixes{}}$) of the reproducible tests
  \item $\mathbf{51.8\%}$ of the fixes (i.e., $\mathbf{\RQfirstnumAccepted{}}$ fixes) were accepted by developers and successfully landed.
\end{itemize}

Table~\ref{tab:category} shows the breakdown of accepted fixes by root cause category\XSpace{, which we discuss below}.
Standard benchmarks are limited to few simple categories, while our industrial setting challenges us with many more complex categories. \flakyguard{} addresses a broader range of flakiness types compared to existing automated tools.

\begin{table}[h!]
\caption{Breakdown of the accepted fixes}
\label{tab:category}
\centering
\footnotesize
\begin{tabular}{lrr}
\toprule
\multirow{2}{*}{\textbf{Category}} & \multicolumn{2}{c}{\textbf{Frequency}} \\
                                   & \textbf{Count} & \textbf{Percentage} \\
\midrule
Schedule randomness & 72 & 37\% \\
Random iteration of unordered collections & 65 & 33\% \\
Timestamp discrepancy & 24 & 12\% \\
State pollution & 16 & 8\% \\
Time-dependent flakiness & 13 & 7\% \\
Others & 7 & 3\% \\
\bottomrule
\end{tabular}
\end{table}



\noindent
\textbf{Schedule randomness.} Schedule randomness is the most common reason for flaky tests.  Go provides extensive support for concurrency, which introduces inherent non-determinism due to thread scheduling. We observe several examples of flaky tests in this category similar to those described in prior work, such as due to asynchronous waits~\cite{LuoETAL2014FSE, RahmanAndShi2024ICSE}.

We show a Go-specific example in Listing~\ref{casestudy:different-path}.  The test function executes \texttt{emitLoop}, which has a \texttt{select} clause. In Go, the language construct \texttt{select} is non-deterministic: when multiple cases are ready, the runtime picks one randomly. In this example, if the runtime picks the first case, it executes only one iteration of the loop. Otherwise, it executes two iterations of the loop, which will fail at the \texttt{Get} mock call, since it is set up to run only once. \flakyguard{} proposed a fix to increase the \texttt{timerPeriod} so that \texttt{ticker.C} becomes ready later than \texttt{ctx.Done()}, ensuring the first case is reliably selected. Developers accepted this fix.


\definecolor{red}{RGB}{200, 0, 0}
\definecolor{darkgreen}{RGB}{0, 128, 0}

\begin{figure}[h]
    \centering
    \begin{lstlisting}[
        frame=single,
        basicstyle=\ttfamily\small,
        xleftmargin=1em,
        xrightmargin=1em,
        breaklines=true,
        linewidth=0.95\linewidth,
        language=CustomGo,
        label={casestudy:different-path},
        caption={A Go-specific flaky test due to the schedule randomness and the select construct.},
        captionpos=b
    ]
// metrics_test.go
t.Run("FailoverCompleted", func(t *testing.T) {
    emitted := runTest(t,
        func(mockK8sClient *clientmock.MockClient) {
            mockK8sClient.EXPECT().
                Get(...).
                DoAndReturn(...)
        },
        func(e *Emitter, ctx context.Context) {
@\rminus@       @\textcolor{red}{e.emit(ctx, time.Millisecond)}@
@\gplus@       @\textcolor{darkgreen}{e.emit(ctx, 200 * time.Millisecond)}@
        },
    )
// metrics.go
func (emitter *Emitter) emit(ctx context.Context, timerPeriod time.Duration) {
   ...
   emitLoop(ctx, timerPeriod)
   ...
}
func (emitter *Emitter) emitLoop(ctx context.Context, timerPeriod time.Duration) {
    ticker := time.NewTicker(timerPeriod)
    defer ticker.Stop()
    for {
        err := emitter.K8sClient.Get(...)

        select {
        case <-ctx.Done():
            return
        case <-ticker.C:
        }
    }
}
    \end{lstlisting}
\end{figure}


We observed around 45\% of the flaky tests with schedule randomness are due to use of \texttt{select} like the one shown above. Other reasons for schedule randomness flaky tests include asynchronous wait (42\%) and atomicity violations, e.g., threads interleave in a way that causes undesired program state (13\%). It is important to note that the LLM needed context from the production code, not just the test code, to determine a fix and produce the reasonable root cause explanation.

\noindent
\textbf{Random iteration of unordered collections.} These flaky tests stem from iterating over unordered data structures, such as Go maps, which do not guarantee consistent traversal order across executions, but the tests implicitly assume a fixed element order~\cite{ShiETAL2016ICST}. Typical mitigation strategies include using relaxed assertions like {\tt ElementsMatch(...)} or explicitly sorting elements prior to comparison. In some complex cases, the results are serialized into strings in the assertion, for which \flakyguard{} deserializes them first.

\begin{figure}[h]
    \centering
    \begin{lstlisting}[
        frame=single,
        basicstyle=\ttfamily\small,
        xleftmargin=1em,
        xrightmargin=1em,
        breaklines=true,
        linewidth=0.95\linewidth,
        language=CustomGo,
        label={casestudy:map-order},
        caption={A flaky test due to non-deterministic map iteration order that leads to an unexpected mock input.},
        captionpos=b
    ]
// decider_test.go
expected := "-9..-1=node/backup,node/backup,node/backup\n-1..10=node/backup,node/backup,node/backup\n"

mockStorage.EXPECT().Write(
    gomock.Any(),
@\rminus@   @\textcolor{red}{gomock.Eq(strings.NewReader(expected))}@
@\gplus@   @\textcolor{darkgreen}{hdfsContentMatcher\{expected, expected2\}}@
).Return(nil)

// production code
func format(input map[string][]string) string {
    var sb strings.Builder
    for token, nodes := range input {
        sb.WriteString(token)
        sb.WriteString("=")
        sb.WriteString(strings.Join(nodes, ","))
        sb.WriteString("\n")
    }
    return sb.String()
}
    \end{lstlisting}
\end{figure}

Listing~\ref{casestudy:map-order} shows an example of such a flaky test. The test sets up the mock so that it accepts an expected input string. However, in production code, the input string is computed from iterating over a map, as shown in function \texttt{format} embedded inside some deep function call chain, which may return different input strings depending on the iteration order. If the input string is not expected, the mock would not be called, thereby leading to an error. \flakyguard{} addresses this problem by providing customized matcher logic for the mock setup that accepts alternative input strings (\texttt{expected2} is omitted for simplicity).




\noindent
\textbf{Timestamp discrepancy.}
If a data structure references a timestamp field transitively, that field can get different {\tt time.Now()} values at different sites, e.g., at the expected value initialization vs the actual initialization. \flakyguard{} typically fixes this type of flaky test by setting the timestamp fields to a constant value or using a comparison logic that ignores those fields. We find that \flakyguard{} needs to extract context from the production code so it can provide the detailed root cause explanation about where the timestamp difference comes from.





\noindent
\textbf{State pollution.}
When multiple tests run in parallel, the state can be modified by other tests~\cite{CandidoETAL2017ASE, EderAndWinter2024ASE}. The pollution can come from shared global state, file system, or databases. \flakyguard{} generates fixes for flaky tests whose pollution comes from all of these sources.

\begin{figure}
    \begin{lstlisting}[frame=single, basicstyle=\ttfamily, xleftmargin=0.1em, xrightmargin=0.1em, breaklines=true, linewidth=1\linewidth, language=CustomGo,label={fig:flaky-env-baggage}, caption=A flaky test due to the pollution., captionpos=b]
// test code:
@\rminus@@\textcolor{red}{    os.Setenv("APP\_ENV", tc.envVar)}@
@\gplus@@\textcolor{darkgreen}{    t.Setenv("APP\_ENV", tc.envVar)}@
populateContextMetadata(ctx)

assert.Equal(t, expectedValue, getSpanBaggage(ctx))

// production code:
func populateContextMetadata(ctx context.Context) {
    span := opentracing.SpanFromContext(ctx)
    group := "default"
    if env := os.Getenv("APP_ENV"); env != "" {
        group = "group_" + strings.ToLower(env)
    }
    span.SetBaggageItem(group)
}
    \end{lstlisting}
\end{figure}

Listing~\ref{fig:flaky-env-baggage} shows an example of such a flaky test. Each test sets the \texttt{APP\_ENV} environment variable (e.g., \texttt{"Staging"} or \texttt{"Production"}) and expects a specific baggage value like \texttt{"group\_staging"} or \texttt{"group\_production"} to be set on the span in the context. However, since environment variables are global to the process, concurrent tests may overwrite each others’ \texttt{APP\_ENV} values. As a result, one test may see baggage derived from another test’s environment setting, causing unexpected assertion failures. The sharing of data through these environment variables leads to nondeterministic behavior when tests are run in parallel.
\flakyguard{} fixed this test by using the \texttt{t.Setenv()} API, following best practice~\cite{tsetenv}, which ensures each test sets and restores the environment properly, and it disallows the tests to run in parallel~\cite{tsetenv}.

%




\noindent
\textbf{Time-dependent flakiness.} The flaky tests in this category depend heavily on the wall clock time.
Listing~\ref{casestudy:time-dependent} shows an example flaky test where there is a deep call chain, i.e., \texttt{GetSupplyChanges} $\rightarrow$ \texttt{validateTimeRange} $\rightarrow$ \texttt{validateTimes} $\rightarrow$ \texttt{validateAge}, which leads to the leaf node function \texttt{validateAge} that explains the root cause. The function computes a new cutoff, in the same way as the test computes the global variable \texttt{cutoff}. However, the two \texttt{time.Now()} calls return different values at runtime. Depending on whether they are truncated to the same second interval, the branch condition at line~\ref{line:time_dependent:branch} may be true or false, thereby making the function return the error non-deterministically.

\flakyguard{} fixes the test by shifting \texttt{beginTime} by 1 second to offset the time elapsed between the test’s start and the invocation of \texttt{validateAge}, so that \texttt{beginTime} is not before the dynamically computed \texttt{cutoff}, avoiding false failures due to the time-dependent flakiness. Such flakiness is hard to fix or root cause without the production code context.


\begin{figure}[h]
    \centering
    \begin{lstlisting}[
        frame=single,
        basicstyle=\ttfamily\small,
        xleftmargin=1em,
        xrightmargin=1em,
        breaklines=true,
        linewidth=0.95\linewidth,
        language=CustomGo,
        label={casestudy:time-dependent},
        caption={A time-dependent flaky test},
        captionpos=b
    ]
// handler_test.go
var cutoff = time.Now().UTC().AddDate(0, -Age, 0).Truncate(time.Second)
func TestGetSupplyChanges(t *testing.T) {
@\rminus@   @\textcolor{red}{beginTime := common.TimeToMS(cutoff)}@
@\gplus@   @\textcolor{darkgreen}{beginTime := common.TimeToMS(cutoff.Add(1 * time.Second))}@
  
    request := &GetSupplyChangesRequest{
        StartTime: beginTime,
        ...
    }
    ...
    _, err := handler.GetSupplyChanges(ctx, request)
    assert.NoError(t, err)
}
// a deep call chain leading here.
func validateAge(...) {
    cutoff := time.Now().UTC().AddDate(0, -Age, 0).Truncate(time.Second)
@\label{line:time_dependent:branch}@    if beginTime.Before(cutoff) {
       ...
       return err
    }
    return nil
}
    \end{lstlisting}
\end{figure}

\noindent
\textbf{Others.}
There are other miscellaneous reasons for flaky tests that \flakyguard{} can fix. One example test draws an insufficient number of samples before asserting over statistical properties such as variance. \flakyguard{} fixes the test by having it draw more samples.
Another example test builds random inputs, where some of the inputs are invalid and break assumptions in the production code. \flakyguard{} generated a fix that restricts the space of random inputs to comply with the assumptions.

\noindent
\textbf{Analysis of the fixes that were not accepted.}
\RQfirstrejectInRight{} of the proposed fixes were not accepted by the developers. One major reason is that the generated fix did not reuse the test helper APIs. Some organizations build such APIs to enforce best practices to ease testing. The LLM we use is not aware of these custom APIs and hence does not use them. Interestingly, through our context collection, the LLM can read the API implementations and copy them over into the fixes. We plan to investigate how to guide LLMs to reuse the APIs directly, which needs to resolve the Bazel build dependencies~\cite{bazel_dependencies}.

Another major reason is that each flaky test may have multiple fixes, where the easier ones are more likely to be produced and sent to developers, but the more complex fixes are more proper and what developers want. In Listing~\ref{casestudy:runningexample}, an easy fix would be to relax the mock so that it can be called maximally once. However, developers expressed to us that they do not want such fixes.


Beyond these reasons, a handful of fixes changed the test semantics, e.g., removing some assertions, adding irrelevant test assertions.
We also saw fixes got rejected because of the readability issues, e.g., the fixes are too long or too complex.

Last but not least, we observed that developers manually fixed some flaky tests in parallel with \flakyguard{}'s fixing pipeline, and they prefer to submit their own fixes.

\subsection{RQ2: \RQTwo{}}~\label{sec:rq2}

We conducted an ablation study to study the contribution of each component of \flakyguard{}. To conduct a controlled experiment (used hereafter up to Section~\ref{sec:rq4}), we selected a single commit of the monorepo (at the time we started the experiment) and performed the evaluations against the \RQSecondnumOfReproducibleTests{} flaky tests  that are reproducible in this commit.
 We used \gptmodel~\citep{o1} and set a two-hour time limit for each test case to fix.

For this ablation study, we did not consider developer feedback, as it would have
introduced extensive back-and-forth processes beyond the scope of the automated setup. Instead, we relied on test validation and spot checking to achieve large-scale results. Although this approach may over-count the impact of certain components, it ensures consistent evaluation criteria across all experimental configurations, thus maintaining comparative validity.


We break the evaluation of this research question into two parts:
\begin{description}
    \item[RQ2.1] How effective are different context collection strategies at fixing the flaky tests?
    \item[RQ2.2] How effective is the test simplification logic at fixing the flaky tests?
\end{description}

\noindent
\textbf{How effective are different context collection strategies at fixing flaky tests?} Table~\ref{tab:level} presents an ablation study addressing RQ2.1. We evaluate four configurations: (1)~\textbf{\NOPROD{}}, which is our re-implementation of FlakyDoctor~\cite{flakydoctor} that provides only the test code context; (2)~\textbf{\BFSRepograph{}}, which performs BFS traversal of static RepoGraphs~\cite{repograph}; (3)~\textbf{\BFSDCG{}}, which performs BFS traversal of dynamic call graphs; and (4)~\textbf{\flakyguard{}}, which applies selective traversal of dynamic call graphs (Section~\ref{sec:graph_traversal}). Note that \BFSRepograph{} serves as another baseline approach similar to FlakyDoctor and does not strictly belong to the ablation study of our technique's components. Configurations (2)-(4) collect relevant production code context starting from the test. Furthermore, for all four configurations, we applied the test simplification techniques described in Section~\ref{sec:simplification_patching} to help them deal with the challenges imposed by the table-driven testing practice.




\NOPROD{} generates \numFixedNoProd{} fixes, while \BFSRepograph{} generates \numFixedBFSRepograph{} fixes. The improvement over \NOPROD{}  demonstrates the usefulness of the production code context. On the other hand, \BFSRepograph{} does not significantly outperform \NOPROD{}. One contributing factor is that the code context, i.e., the transitive closure over the static graph, can get so big that it confuses the LLM.

\BFSDCG{} generates \numFixedBFSDCG{} fixes. The improvement over \BFSRepograph{} highlights the value of the dynamic call graph, which helps produce more focused and precise context by excluding the unexecuted branches and precisely resolving the dynamic call dispatch. In addition, we observed cases where \BFSDCG{}  collects the
relevant context hidden behind the reflection calls or anonymous functions, while \BFSRepograph{} is unable to.

We would like to mention that the performance gap between \BFSDCG{} and \BFSRepograph{} would be substantially larger without test simplification. Most test functions follow table-driven testing practices~\cite{tabledriventests}, containing typically 10+ test cases. Without simplification, \BFSRepograph{} presents the complete test function to the LLM, causing it to often focus on unrelated test cases (in around 60\% of cases).

\flakyguard{} can successfully generate \numFixedSmartBFS{} fixes, which outperforms \BFSDCG{} by 20.5\%, thanks to the selective context collection.
Dynamic call graphs can be substantial, with the node count reaching hundreds  and the graph depth extending to seven levels in extreme cases.  In the cases that can be fixed by \flakyguard{} but not by \BFSDCG{}, \flakyguard{} helped greatly in selecting a smaller set of nodes as context and letting the LLM focus on a smaller relevant scope.

To quantify the effectiveness of our core contribution of intelligent context pruning, we analyzed the pruning mechanism in detail. In Listing~\ref{casestudy:runningexample}, \flakyguard{} selected 27 nodes from the DCG with 264 nodes, then further limited it to 5 nodes via the global filtering. This dramatic reduction from hundreds of nodes to approximately 5 maximally relevant functions enables the LLM to focus on semantically important code while eliminating extraneous information. Manual inspection of cases fixed by \BFSDCG{} revealed that the root cause nodes are selected by \flakyguard{} in all but two cases, confirming the precision of our selection mechanism.

\begin{table}[t]
\caption{Comparison of different context collection strategies.}
\label{tab:level}
  \centering
\footnotesize
\centering
\begin{tabular}{lcc}
\toprule
\textbf{Approach} & \textbf{Fixed Tests} & \textbf{Success Rate (\%)} \\
\midrule
FlakyDoctor & 134 & 45.42 \\
BFSRepoGraph & 149 & 50.51 \\
BFSDCG & 157 & 53.22 \\
FlakyGuard & 194 & 65.76 \\
\bottomrule
\end{tabular}
\end{table}


\noindent
\textbf{How effective is the test simplification logic at fixing the flaky tests?} We conduct an ablation study to assess the effectiveness of the test simplification (Section~\ref{sec:simplification_patching}).
When comparing \flakyguard{} before and after applying simplification, the generated fixes increase from \numFixedSimplification{} to \numFixedSmartBFS{}, demonstrating that simplification significantly improves performance in addressing the complexity introduced by table-driven testing practices, which result in test functions with multiple similar-looking tests. Sometimes, the tests only differ with a few words in the description and the test body differs only on few arguments, which imposes a great challenge to the LLM and often misleads it to look into some other irrelevant tests. The test simplification helps the LLM focus only on the test case of interest and the functions actually called by it.

\subsection{RQ3: \RQThree{}}\label{sec:rq3}
\begin{table}[t]
\caption{Comparison of \flakyguard{} with \Agentless{} and \AutoCodeRover{}.}
\label{tab:different_approaches}
  \centering
\footnotesize
\centering
\begin{tabular}{lcc}
\toprule
\textbf{Approach} & \textbf{Fixed Tests} & \textbf{Success Rate (\%)} \\
\midrule
Agentless+RepoGraph & 159 & 53.90 \\
AutoCodeRover & 158 & 53.56 \\
FlakyGuard & 194 & 65.76 \\
\bottomrule
\end{tabular}
\end{table}

We compare \flakyguard{} with \Agentless{}~\cite{agentless,repograph} and \AutoCodeRover{}~\cite{autocoderover}.
\Agentless{} and \AutoCodeRover{} are the state-of-the-art approaches designed to fix Python bugs (evaluated on SWE-Bench).
We extracted the context collection components from the publicly available code, which we believe is the major factor for the success of the LLM.

For \Agentless{}, we combine Agentless~\cite{agentless} and RepoGraph~\cite{repograph} with modifications to the RepoGraph component. Since the original approach searches for callers to identify the failure-inducing usage at the caller side, we adapted it to search for callees to understand implementation details that cause flakiness. We use one-hop traversal following the default configuration~\cite{repograph, repograph_github}. We exclude the embedding-based file search from Agentless~\cite{agentless} as it requires natural language descriptions from developers, which are not available in our problem setting.

For \AutoCodeRover{}, we directly use the existing technique without additional customizations, like we did for \Agentless{}. It iteratively searches for relevant functions starting from the flaky test and stops when it believes it has found sufficient context for fixing.

We enhanced \Agentless{} and \AutoCodeRover{} by applying the techniques in Section~\ref{sec:simplification_patching} to tackle the challenges imposed by the table-driven testing in order to produce meaningful results. Without these enhancements, they would constantly get confused working on some irrelevant test cases or encounter patching errors due to the matching ambiguity.

Table~\ref{tab:different_approaches} shows the  results of the comparison. \Agentless{} generates  \numFixedAgentless{} fixes, each taking \agentlesstime{} seconds on average. \AutoCodeRover{} generates \numFixedAutoCodeRover{} fixes, each taking \autocoderovertime{} seconds on average. In contrast, \flakyguard{} fixes \numFixedSmartBFS{}, each taking  \smartbfstime{} seconds on average. \flakyguard{} outperforms \Agentless{} or \AutoCodeRover{} by 22\%, which quantitatively confirms the effectiveness of our technique.
In terms of overall efficiency, although \flakyguard{} incurs additional overhead for code instrumentation and runtime call graph collection, we observed \flakyguard{} is usually more efficient in the LLM reasoning and fixing phases than the other two because (1)~with shorter context, each LLM response is faster; and (2)~with more relevant context, the LLM succeeds within fewer iterations. The initial instrumentation overhead is justified by the subsequent reduction in patch generation time.

\begin{figure}[t]
    \centering
    \includegraphics[width=0.8\linewidth]{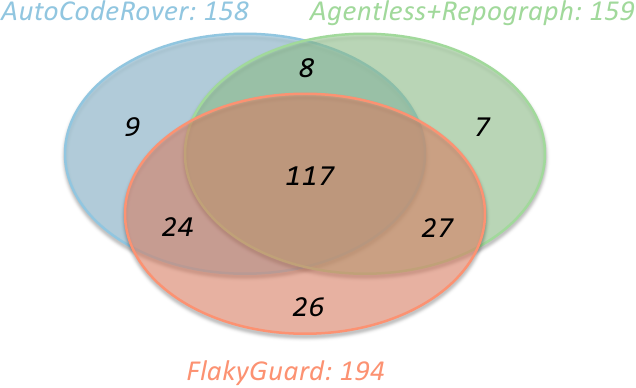}
    \caption{Venn diagram for the fixes generated by different approaches}
    \label{fig:venn}
\end{figure}

Figure~\ref{fig:venn} shows a Venn diagram illustrating the overlap of the fixes produced by each approach. No approach's fixed tests are a subset of any other, suggesting that each one has its pros and cons.
We manually inspected the tests that each approach cannot fix while the others can fix, during which we paid special attention to the contexts collected.

\noindent
\textbf{Flaky tests that \Agentless{} cannot fix.}
We manually analyzed \numFixedBySBFSNotAgent{} flaky tests that were fixed by \flakyguard{} but not by \Agentless{}, identifying three main reasons \Agentless{} could not fix tests. First, due to its lack of dynamic information, \Agentless{} often picks the wrong function or includes code branches not executed by the test, resulting in large and noisy prompts that confuse the LLM (13 tests). Second, \Agentless{} relies on static name-based analysis, which causes it to miss relevant anonymous functions or functions invoked via reflection (9 cases). Third, the Repograph in \Agentless{} employs a shallow graph traversal strategy: it uses a one-hop call graph by default~\cite{repograph} to avoid exponential context growth, which leads to missing the root causes deeply nested in the leaf nodes (9 cases).

\noindent
\textbf{Flaky tests that \AutoCodeRover{} cannot fix.}
\AutoCodeRover{} failed to fix \numFixedBySBFSNotAutocoderover{} flaky tests due to three main reasons. First, \AutoCodeRover{} often terminates early after identifying a potential bug location, but in some cases, additional surrounding context is required for successful fixing (15 cases). Second, like \Agentless{}, \AutoCodeRover{} can include excessive irrelevant context  (11 cases). Third, due to the limitations of static analysis, \AutoCodeRover{} also misses the relevant anonymous functions or reflection-based functions (9 cases).

\noindent
\textbf{Flaky tests that \flakyguard{} cannot fix.}
Among the flaky tests we analyzed, 24 were fixed by either \Agentless{} or \AutoCodeRover{} but not by \flakyguard{}. These failures fall into three categories. First, \flakyguard{} either missed relevant functions or included irrelevant ones, despite traversing the dynamic call graph (9 cases). However, compared to other approaches, the graph-guided traversal greatly reduced the number of such cases overall. Second, the global filtering \flakyguard{} uses during postprocessing removed relevant functions, causing the LLM to fail in the fixing (8 cases). Third, \flakyguard{} provided similar context as the other tools did but failed to generate a fix, which may be due to differences in the function ordering within the prompt or inherent randomness in using the LLM (7 cases).

\subsection{RQ4: \RQFour{}}\label{sec:rq4}

\begin{figure}[t]
    \centering
    \includegraphics[width=0.6\linewidth]{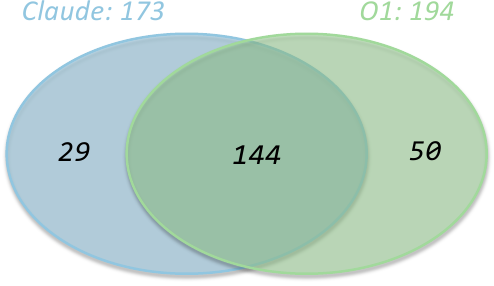}
    \caption{Venn diagram for the fixes generated with different models}
    \label{fig:venn2}
\end{figure}

We compared \gptmodel~\citep{o1} and \claudemodel~\cite{claude} in this study. Figure~\ref{fig:venn2} shows a Venn diagram illustrating the overlap of the fixes produced with each model. No model's fixed tests are a subset of the other, suggesting that each one has its pros and cons.

\noindent
\textbf{Flaky tests that \claudemodel{} cannot fix.}
\claudemodel{} failed to fix \numFixedByGPTNotClaude{} flaky tests that \gptmodel{} successfully addressed.
In 38 tests, \claudemodel{} selected different contexts during the graph traversal, which yielded incorrect fixes. In 12 tests, \claudemodel{} selected the same context as \gptmodel{} but still produced incorrect fixes. In 4 out of these 12 cases, \claudemodel{} derived the wrong high-level root causes. Due to the black box nature, it is beyond our capability to explain what happened.
When we manually inspected these cases, we found that the root causes given by \claudemodel{} were also plausible, but the fixes failed the automatic validations.

\noindent
\textbf{Flaky tests that \gptmodel{} cannot fix}
\gptmodel{} failed to fix \numFixedByClaudeNotGPT{} flaky tests that \claudemodel{} successfully addressed.
In 22 tests, the two models selected different context during graph traversal, which yielded incorrect fixes. In 7 tests, \gptmodel{} shared the same context with \claudemodel{}, yet still produced incorrect fixes. In 3 out of these 7 cases, \gptmodel{} derived the wrong high-level root causes. Like the previous case of \claudemodel{}, \gptmodel{} gave plausible but subtly wrong explanations. The fixes did not pass the validations.

\subsection{RQ5: \RQFive{}}~\label{sec:rq5}
\begin{table}
    \centering
    \scriptsize
    \caption{Survey Results on Developers' Perceptions of \flakyguard{}}
    \label{tab:survey_results}
        \centering
        \begin{tabular}{ll}
            \toprule
        \midrule
        \multicolumn{2}{c}{\textbf{Usefulness of the root cause explanation}} \\
        \midrule
        \textbf{Usefulness} & \textbf{Count(\%)} \\
        \midrule
        Uesful                & 19  (100\%) \\
        \midrule
        \multicolumn{2}{c}{\textbf{Quality and Complexity of flaky tests fixed by \flakyguard{}}} \\
        \midrule
        \textbf{Metric} & \textbf{Average Rating (stddev)} \\
        \midrule
        Quality of Fixes (1-5)               & 4.42 $\pm$0.77 \\
        Complexity of Root Causes (1-5)       & 2.73 $\pm$1.10 \\
        \midrule
        \multicolumn{2}{c}{\textbf{Estimated Time Saved by \flakyguard{}}} \\
        \midrule
        \textbf{Time Saved} & \textbf{Count(\%)} \\
        \midrule
        1 to 8 hours                & 3  (15.8\%) \\
        less than 1 day             & 8  (42.1\%) \\
        1 to 2 days                 & 6  (31.6\%) \\
        2 to 4 days                 & 2  (10.5\%)  \\
        \bottomrule
    \end{tabular}
\end{table}
We sent out surveys to developers to understand their perceptions to \flakyguard{}, receiving 19 responses. Table~\ref{tab:survey_results} shows the results. All 19 respondents (100\%) reported that \flakyguard{}’s root‐cause explanations are useful.

On a five‐point scale, developers rated the quality of the fixes produced by \flakyguard{} very highly (mean 4.42, standard deviation 0.77), while the complexity of the underlying root causes was judged to be moderate (mean 2.73, standard deviation 1.10).

In terms of time savings per individual flaky test case, 3 participants (15.8\%) estimated saving between one and eight hours, 8 (42.1\%) estimated saving less than one day, 6 (31.6\%) estimated saving one to two days, and 2 (10.5\%) estimated saving two to four days when resolving flaky tests with \flakyguard{}. Together, these results suggest that not only do developers find \flakyguard{}'s diagnostic information valuable, but they also perceive \flakyguard{}'s solutions to be of high quality, as well as saving them time.

\subsection{Limitations and Threat to Validity}

LLM reasoning and fixing are naturally affected by non-determinism, where the LLM may fail to repair a flaky test in one attempt but succeed in a later attempt even with identical context.
LLM non-determinism represents a potential threat to evaluation validity~\cite{ouyang2025empirical}. To explicitly mitigate this issue, our experimental design incorporates multiple retry mechanisms with $M=3$ context collection attempts, $P=2$ parallel reasoning paths, and $N=3$ fix generation attempts per path (as detailed in Section~\ref{sec:setup}), providing 18 total repair opportunities per test case. These retry mechanisms directly help reduce the impact of non-deterministic LLM behavior by providing multiple chances for successful repair. Additionally, our concrete and detailed prompts are designed to minimize output variability, which according to Ouyang et al.~\cite{ouyang2025empirical} can significantly reduce non-determinism in LLM responses.

\subsection{Discussion}
\flakyguard{}'s methodology is generally applicable to other languages. We implemented a version for Java and confirmed it works on Java flaky tests at \company{}. The approach requires minimal language-specific adaptation, where only the flakiness pattern prompts need to be adjusted. The patterns are shown in Table~\ref{tab:category}, where each pattern is described by a single sentence. One can add prompts for other languages easily.

We chose Go as our evaluation target because it exhibits a rich variety of flakiness patterns in our industrial setting. Our dataset spans more than 100 projects with diverse business logic, revealing flakiness types not extensively studied in prior work. This comprehensive coverage motivated the development of \flakyguard{}.

The techniques we employ are standard software engineering practices. Table-driven testing follows official Go guidelines, and after simplification, these tests become conventional unit tests. The flakiness patterns we address represent universal challenges in concurrent software development across languages and organizations.

The main change we had to make was on the dynamic call graph collection part, which needs instrumentation, which we can do with most modern compilers or even with Tree-sitter~\cite{treesitter} at the AST level, since we only need to insert one API call at the method entry. When dynamic call graphs are unavailable, one can still apply the LLM-guided graph exploration on the static call graph or other scalable graph format~\cite{scip}. In addition, we made new implementations of the test simplification and patch transplantation logic.

Our open-source implementation includes dynamic call graph collection mechanisms, LLM-guided context selection algorithms, and automated patch generation capabilities. The implementation and evaluation artifacts are available at: \url{https://sites.google.com/view/flakyguard}.

\section{Related Work}
\label{sec:rel}


Luo~\etal{} conducted the first empirical study on flaky tests in open-source
projects, categorizing the reasons they are flaky as well as how developers
fixed them~\cite{LuoETAL2014FSE}. Following this study, researchers
developed techniques to detect and repair specific types of flaky tests automatically. 
For example, to detect order-dependent flaky tests, which are tests whose
pass/fail outcome depend on which tests run before it due to pollution in shared
state, researchers proposed techniques to run tests in different
orders~\cite{ZhangETAL2014ISSTA, LamETAL2019ICST, LiAndShi2022ISSTA,
LiETAL2023ISSTA, WeiETAL2021TACAS, WeiETAL2022ICSE, WangETAL2022ICSEDemo,
HashemiETAL2025ICST} or to track shared state between tests to see what could be
polluted~\cite{GyoriETAL2015ISSTA, GambiETAL2018ICST}. Researchers later
developed techniques to automatically repair these flaky tests by identifying
code within the application that can be used as a patch into the test code to
reset the shared state~\cite{ifixflakies, odrepair}. Shi~\etal{}~\cite{ShiETAL2016ICST}
developed NonDex to detect tests that fail due to assuming an ordering on
unordered collections by exploring how tests behave when controlling over the
iteration order on those collections~\cite{ShiETAL2016ICST,
GyoriETAL2016FSEDemo}. Zhang~\etal{} later developed DexFix to repair these
flaky tests~\cite{dexfix}. For timing-dependent flaky tests,
researchers developed techniques that efficiently explore where to insert delays
that can reliably reproduce their failures~\cite{RahmanETAL2024ICSTFlakeRake,
LeesatapornwongsaETAL2022FSE}. Techniques for repairing timing-dependent flaky
tests include adjusting existing wait times in the code~\cite{LamETAL2020ICSE,
YuETAL2025TOSEM} or synchronizing the critical
points~\cite{RahmanAndShi2024ICSE}. We found examples of all of these flaky
tests in \company{}, and we leverage our LLM-based approach to automatically
repair them.

Recently, researchers are using LLMs to solve flaky test-related problems, such
as predicting which tests are flaky~\cite{FatimaETAL2023TSE,
RahmanETAL2024ICSTFlakyQ, RahmanETAL2025OOPSLA} or debugging flaky tests~\cite{AkilETAL2023AST,
RahmanETAL2024ICSTFlakyQ, RahmanETAL2025ICSE}. Our work is most similar to
FlakyDoctor~\cite{flakydoctor}, an LLM-based approach to automatically repair
flaky tests. FlakyDoctor is effective at repairing order-dependent flaky tests
and flaky tests that assume deterministic ordering of unordered collections,
repairing more tests than non-LLM approaches~\cite{ifixflakies,
odrepair, dexfix}. Our approach also collects context from test code and executions to prompt an LLM for a
solution. However, we leverage a more generic way of collecting context based on
searching over a dynamic call graph, allowing us to generally handle more types
of flaky tests, including those that FlakyDoctor handles. Fatima~\etal{}
evaluated the effectiveness of LLM-based flaky test repair if the model also
received the fix category along with some examples, showing this extra
information helps generate better fixes~\cite{flakyfix}. We also prompt the LLM by summarizing the pattern from previous examples to the model to help predict the category and generate a fix, along with extra context.

Recent work leveraged LLMs with different context extraction methods for various software engineering tasks. Agentless~\cite{agentless} uses hierarchical localization from repository overview to specific code elements to obtain the context. AutoCodeRover~\cite{autocoderover} combines LLM with program analysis, using stratified retrieval and iterative API calls to gather relevant code snippets. RepoGraph~\cite{repograph} creates a code graph using Tree-sitter, enabling k-hop ego-graph retrieval for related semantic context. We adapt all three of these past approaches for our goal of repairing flaky tests, to use as comparison against \flakyguard{}. \flakyguard{} leverages a dynamic call graph and a selective LLM-based graph search to more effectively and efficiently repair flaky tests.


\section{Conclusions}
\label{sec:concl}

We presented \flakyguard{}, a novel approach that addresses the context problem in LLM-based flaky test repair through selective exploration of dynamic call graphs. Our evaluation on real-world industrial flaky tests shows that \flakyguard{} achieves high developer acceptance, outperforming existing methods by at least 22\%. The approach effectively balances providing sufficient context for accurate repairs while avoiding information overload, and developers find the root cause explanations universally useful. \flakyguard{} represents a significant advancement in automated flaky test repair for industrial software development.

In addition, \flakyguard{} has been deployed as a fully autonomous system at \company{}, integrating with the existing ticketing system to automatically process flaky test reports and deliver fixes to developers daily.

\bibliographystyle{IEEEtran}
\bibliography{references}
\end{document}